# Unconsented Sensing:

# A Sociotechnical Governance Framework for 6G ISAC

Anass Sedrati
*Department of Communication Systems*
*KTH Royal Institute of Technology*
Stockholm, Sweden
anass@kth.se

***Abstract*—The forthcoming deployment of 6G Integrated Sensing and Communication (ISAC) will transform cellular infrastructure into pervasive, continuous environmental and biometric sensing grids. While current telecom standardization efforts (e.g., 3GPP, ETSI) have formally recognized privacy and trustworthiness as critical pillars for 6G, their proposed mitigations remain overwhelmingly technocentric, relying on cryptographic anonymization and physical layer security. This approach critically underestimates the sociotechnical and legal complexities of the downstream machine learning (ML) models required to interpret raw sensing data, creating a profound collision with existing digital rights legislation. This position paper argues that technical security is insufficient. ISAC trustworthiness must be redefined as mandatory regulatory and sociotechnical compliance. We identify the specific legal friction points between continuous ISAC surveillance and the mandates of emerging global digital rights regimes, using the stringent requirements of the EU AI Act and GDPR as our primary regulatory baselines. To bridge this gap, we propose a governance framework centered on three pillars: Purpose-bound sensing activation, citizen transparency mechanisms, and algorithmic accountability for ISAC-driven ML models. Ultimately, this paper provides a regulatory roadmap to prevent the illegal deployment of 6G sensing infrastructures and ensure they remain viable before physical deployment.**

## I. Introduction

The standardization of 6G will introduce Integrated Sensing and Communication (ISAC) as a native capability, transforming cellular infrastructure from a passive data conduit into an active environmental radar [1], [2]. By multiplexing communication and radar functions over shared hardware and radio frequency (RF) spectra, ISAC enables the continuous monitoring of physical spaces. Interpreting raw signal reflections, such as Channel State Information (CSI) and micro-Doppler shifts, relies on advanced Machine Learning (ML) algorithms [3]. These ML models are deployed to classify and infer sensitive data, ranging from broad environmental mapping to granular Human Activity Recognition (HAR), including presence detection, gesture tracking, and vital sign monitoring [4].

This shift creates a regulatory challenge: How can operators secure meaningful legal consent for an invisible radar spanning the public space? Unlike wearable sensors or cameras, ISAC targets are passive and often non-consensual. Individuals moving through a monitored RF field do not opt-in to transmitting data, yet their physiological and spatial states are continuously inferred.

While standardization bodies start acknowledging these privacy risks, technical mitigations remain inadequate. For instance, 3GPP Release 19 [5] and ETSI GR ISC 004 [6] recognize the theoretical need for sensing consent, transparency, and AI model trustworthiness. However, their approaches still rely heavily on physical layer security, access control authorization, and cryptographic anonymization. While these standards identify the legal mandates of the General Data Protection Regulation (GDPR) and the EU Artificial Intelligence (AI) Act, they treat them as abstract policy gaps rather than engineering issues. Consequently, current standardization frameworks lack a sociotechnical architecture to enforce data minimization, purpose limitation, and algorithmic auditing prior to data generation.

This paper argues that the engineering definition of ISAC trustworthiness must be redefined beyond cryptographic resilience to include regulatory compliance by design. To bridge this critical gap, the remainder of this paper is organized as follows: Section II maps the continuous, ML-driven surveillance capabilities of 6G against the legal friction points of European digital laws. Section III proposes our core contribution: A governance framework centered on three architectural pillars: Purpose-bound sensing activation, citizen transparency mechanisms, and algorithmic accountability. Section IV critically assesses implementation challenges and computational latency trade-offs of this framework, pointing toward future work, before concluding the paper in Section V.

## II. Regulatory Challenges of Continuous Sensing in ISAC

Integrating sensing into 6G shifts telecommunications from a strictly data-transport paradigm to one of continuous environmental data generation. While recent 6G frameworks prioritize physical layer security and encryption [7], [8], they fail to address the legality of initial data generation. As European data protection guidelines establish, post hoc anonymization cannot retroactively legitimize unlawful data collection [9]. When ISAC base stations utilize ML models to interpret RF reflections, they trigger severe regulatory mandates under European digital rights frameworks, most notably the AI Act [10] and the GDPR [11], which both provide rigorous baselines used by global operators.

### A. *The Classification of ISAC Sensing Data Under the EU AI Act*

A core utility of ISAC lies in its ability to detect and classify environmental objects and human presence using RF signals rather than optical cameras [1], [2]. Through the analysis of micro-Doppler signatures and Channel State Information (CSI), edge-deployed ML algorithms can infer human presence, gait, gestures, and physiological metrics such as respiration and heart rates [12], [13], [14], [15]. However, under the recently enacted EU AI Act, translating ambient radio waves into human behavioral or physiological insights exponentially raises the system's regulatory risk classification.

The critical friction point is whether ISAC-driven ML models constitute Biometric Categorization or Real-Time Remote Biometric Identification (RBI). According to the AI Act, systems that categorize natural persons based on their biometric data to deduce or infer protected characteristics, behaviors, or health states are subject to stringent transparency and auditing requirements, frequently falling under Annex III 'High-Risk' classifications [10]. If a 6G base station infers the breathing rate of individuals in a public square or classifies pedestrian density based on localized RF disruptions, the company operating it ceases to be a mere conduit of information and becomes an operator of a high-risk AI system, compelling the operator to comply with several additional requirements.

Moreover, the development of ISAC applications for high-precision gait recognition [12] fundamentally conflicts with RBI prohibitions. The AI Act explicitly prohibits the use of 'real-time' remote biometric identification systems in publicly accessible spaces even for law enforcement purposes, allowing narrow, strictly defined exceptions. If future 6G networks possess native RBI capabilities via continuous RF sensing, their deployment in public spaces risks creating an unlawful, ubiquitous biometric surveillance grid by default. Treating these ML inferences merely as network optimization data ignores that European AI governance protects society structurally. Because ubiquitous sensing operates indiscriminately, individual privacy deficits inherently aggregate into systemic societal risks.

### B. *The GDPR and the Paradox of Continuous Sensing*

The foundational premise of ISAC, which is that the network continuously "senses" to maintain situational awareness, creates a structural incompatibility with the GDPR. Within current technical literature, privacy preservation is predominantly approached as a data security problem. The prevailing assumption is that if ISAC data is cryptographically anonymized prior to transmission from the edge server, privacy is maintained [5], [6]. This technocentric view fundamentally conflicts with GDPR Article 5(1)c: Data Minimization, and Article 5(1)b: Purpose Limitation [11].

The principle of data minimization legally dictates that personal data collection must be strictly limited to what is necessary for a specific, explicitly stated purpose. The notion of a pervasive sensing infrastructure operating continuously fundamentally violates this mandate on two fronts: (1) It indiscriminately captures vastly more spatial and biometric data than any single service could justify, and (2) in an "always-on" deployment, it does so without any predefined, explicit purpose. The legal violation occurs at the exact moment of this unjustified sensing and algorithmic inference, rendering the data collection fundamentally illegal, regardless of whether the resulting data is subsequently encrypted. Post hoc anonymization is merely a data processing technique; it does not retroactively legitimize the initial unlawful data collection.

Moreover, the assertion that ISAC merely processes anonymous kinematic patterns rather than personal identities is legally insufficient. Even if a pedestrian is not a subscriber to the operator or is walking without a device, the unique nature of micro-Doppler signatures allows systems to persistently track them across multiple base stations. When individuals do carry active devices, telecommunications operators possess the inherent capability to cross-correlate these “anonymous” physical signatures with known mobile network trajectories. In either scenario, this capability satisfies the GDPR’s legal threshold for “singling out” an individual, effectively rendering the collected RF signatures personal data.

This spatial and identifying nature of ISAC data exacerbates the issue of lawful consent (GDPR Article 6). In contrast to traditional digital services where explicit consent mechanisms can be readily implemented, individuals navigating physical environments lack practical means to opt out of device-free ambient RF sensing. Analogous to the legal precedents established by the European Data Protection Board (EDPB) regarding continuous video surveillance [16], passive, non-cooperative subjects cannot legally provide implicit consent merely by occupying a public street.

This challenge to traditional consent mechanisms is largely driven by the emerging Sensing as a Service (S2aaS) business model, which financially incentivizes Communication Service Providers (CSPs) to monetize continuous environmental data [17]. This marks a profound legal escalation from 4G and 5G network paradigms, where operators commercialize the aggregated location data of mobile network subscribers, to a 6G paradigm that commodifies the device-free, physical movements of the general public without any direct contractual relationship.

However, deploying this ubiquitous sensing infrastructure introduces severe, unintended consequences to the CSPs' structural threat landscape. By retaining pervasive environmental data without a defined operational necessity, operators inadvertently transition from mere infrastructure conduits to aggregators of highly sensitive behavioral profiles. This architectural paradigm creates centralized repositories of critical vulnerability, exponentially expanding their attack surface against malicious exfiltration and subjecting them to an escalated volume of compulsory governmental access requests, thereby engendering severe operational and legal liabilities.

To render 6G ISAC legally viable, the industry cannot rely on post hoc data obfuscation. Trustworthiness must be structurally designed into the system prior to deployment, mandating a governance framework that strictly limits when, why, and how sensing capabilities are activated.

## III. Proposed Governance Framework for Trustworthy ISAC

Treating data protection solely as a cryptographic problem is legally insufficient. We propose a governance framework embedding regulatory mandates directly into the ISAC architecture. These pillars operate sequentially: establishing a lawful baseline for hardware activation (pillar 1), ensuring societal awareness of surveillance (pillar 2), and mandating algorithmic accountability for data processing (pillar 3). The physical integration of these three pillars into the 6G network topology is illustrated in Figure 1 below.

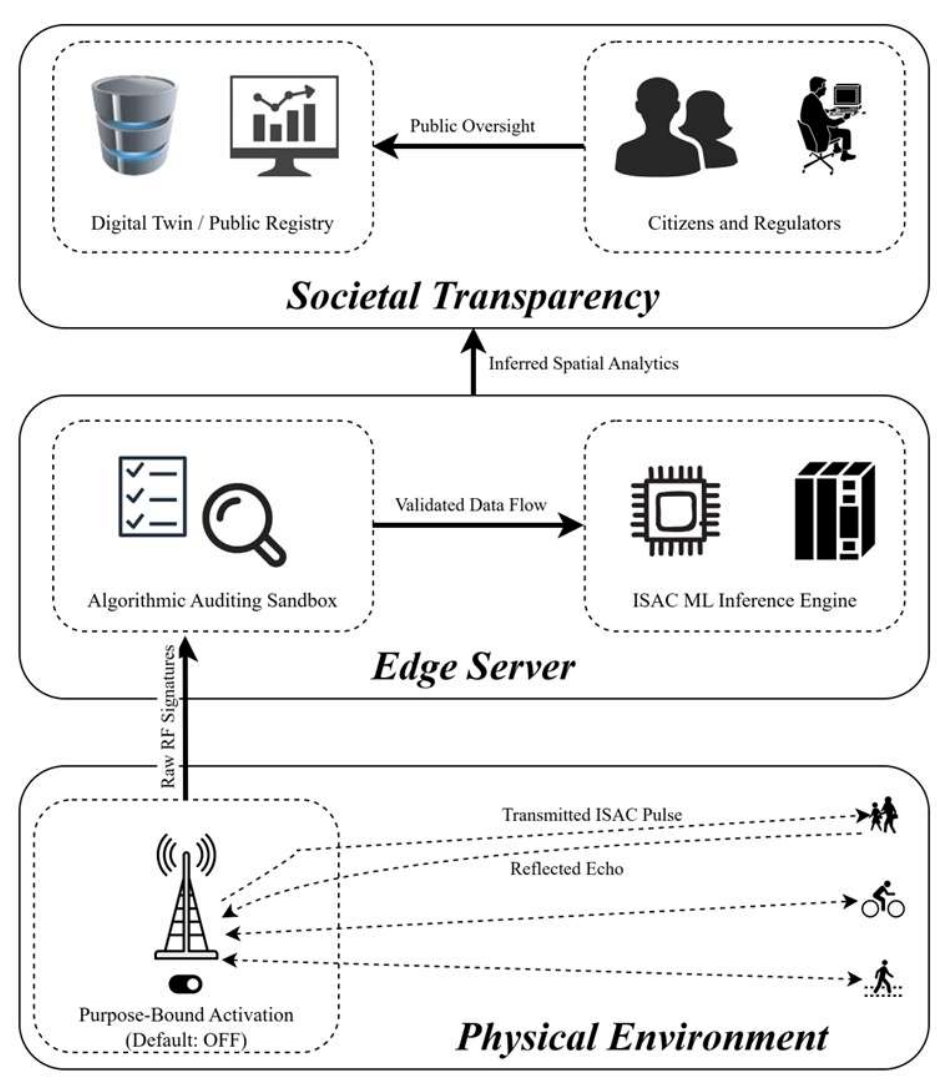

Fig. 1. Proposed Sociotechnical Architecture for Trustworthy 6G ISAC

### A. *Pillar 1: Purpose-Bound Sensing Activation via Hardware Controls*

To align with the GDPR’s fundamental principle of data minimization, current network deployment paradigms must abandon an “always-on” sensing grid. Continuous, device-free ambient RF sensing of public spaces without an explicit, verifiable objective is legally untenable. Consequently, 6G ISAC base stations cannot be designed to leave their sensing capabilities perpetually active merely to accumulate ambient environmental data.

Instead, the infrastructure must enforce purpose-bound sensing. Compliance necessitates demonstrable hardware and software mechanisms, such as reconfigurable RF architectures utilizing spatial null steering to physically mask sensitive zones [18] and software means to prohibit passive sensing [6], ensuring the device-free radar capability remains strictly dormant by default. Under this architecture, the ISAC sensing function acts as an inactive hardware capability that can only be triggered via authenticated, policy-driven API calls.

Crucially, these activation triggers must correspond strictly to specific, legally cleared purposes, balancing rights protection with operational necessity. For example, sensing might be activated in a highly localized sector to assist emergency responders in locating victims during a structural collapse (public safety/disaster response), or briefly toggled on for municipal traffic optimization based on a pre-approved smart city mandate. Once the legally justified task is completed, the system must default back to a dormant state. Failing to enforce this immediate return to dormancy reinstates the continuous data accumulation loop, directly violating the GDPR's mandate for data protection by default (Article 25(2)) and recreating the exact operational liabilities and centralized data vulnerabilities established in the previous section.

## B. *Pillar 2: Dynamic Transparency and Spatial Consent Mechanisms*

The deployment of an invisible, ubiquitous sensing network fundamentally nullifies traditional digital consent models. Because ISAC operates as a device-free radar, citizens cannot "opt-out" by simply refusing to carry a 6G-enabled smartphone or clicking "disagree" on a digital privacy banner. To satisfy the transparency mandates of the GDPR (Articles 12-14), requiring data subjects to be informed of data collection, telecom operators must develop novel, spatially aware signaling mechanisms to communicate ISAC operational states to the public.

Given that physical signage is impractical for dynamic, sector-specific RF beams, we propose the integration of ISAC activation states into municipal "Digital Twins" or real-time public registries [19], [20], or, as a transitional mechanism where such civic infrastructure is absent, public transparency dashboards and open APIs maintained by operators. In this model, whenever a 6G base station activates its sensing capabilities, it must broadcast a standardized transparency beacon to a public-facing geographic information system (GIS). Citizens, regulators, and civil society organizations could access this registry to dynamically verify which specific physical zones are currently subjected to RF sensing, the precise ML inferences being drawn (e.g., traffic density vs. pedestrian tracking), and the lawful basis for that activation. By mapping invisible RF surveillance into a visible, auditable digital interface, this pillar transitions ISAC from an opaque commercial deployment into a transparent civic infrastructure.

## C. *Pillar 3: Algorithmic Auditing of Sensing Models*

The operationalization of ISAC necessitates the deployment of Machine Learning (ML) models at the network edge to classify raw RF reflections, such as micro-Doppler signatures or Channel State Information, into actionable insights. Because these models are tasked with inferring human presence, behavioral states, or physiological metrics, they inherently risk triggering 'Biometric Categorization' or 'High-Risk' classifications under the regulatory scope of the EU Artificial Intelligence Act.

Consequently, the ML models interpreting ISAC radar data must be subjected to rigorous, mandatory auditing for bias, accuracy, and robustness prior to live deployment. For instance, even in scenarios where an operator has established a lawful, purpose-bound mandate for municipal pedestrian detection (as governed by pillars 1 and 2), the underlying edge server model must still undergo strict AI Act auditing. This ensures the system does not generate discriminatory inferences or systematically fail to detect specific demographic groups (children or individuals using wheelchairs) due to biased RF training data.

To operationalize this rigorous auditing, we propose the establishment of standardized regulatory sandbox testbeds where telecom operators must empirically prove their ISAC ML models adhere to the AI Act's Annex III requirements. Beyond initial deployment, continuous post-market monitoring must also be mandated to prevent concept drift, ensuring that ambient environmental changes do not cause the system to misclassify benign human activities as anomalous or threatening behaviors.

Ultimately, implementing this three pillar framework ensures that the forthcoming 6G network transcends pure physical layer optimization, establishing a secure, lawful, and societally trusted infrastructure.

## IV. Implementation Challenges and Future Research Directions

To operationalize this conceptual framework, the three pillars must map directly to Open RAN (O-RAN) and 3GPP architectures. First, purpose-bound activation can be enforced via Service Management and Orchestration (SMO). Here, Non-Real-Time RAN Intelligent Controller (Non-RT RIC) rApps can translate legal mandates into strict hardware activation policies via the O1 interface [21]. For transparency mechanisms, the architecture can leverage the 3GPP Network Exposure Function (NEF) to securely broadcast sensing states to public APIs. Finally, algorithmic auditing maps directly to the 3GPP Network Data Analytics Function (NWDAF). To protect proprietary edge-model weights during these compliance checks, operators can utilize Zero-Knowledge Proofs (ZKPs) for ML [22] and verifiable federated protocols [23], enabling regulators to mathematically verify model fairness and data minimization without accessing the raw sensing data.

While these architectural mappings align 6G ISAC with European legal mandates, their practical deployment introduces non-trivial engineering, administrative, and infrastructural challenges. Transitioning from a theoretical governance model to an operational network standard requires bridging critical gaps between physical layer constraints and real-time compliance mechanisms.

First, mapping pillar 1 to the SMO isolates governance from real-time execution. By processing legally authenticated authorization protocols via the Non-RT RIC rather than the RF front-end, the architecture prevents computational latency from degrading Ultra-Reliable Low-Latency Communications (URLLC) [24]. Furthermore, the administrative overhead required for telecommunications operators to manage and verify dynamic API calls for localized sensing activation necessitates a fundamental restructuring of traditional network orchestration paradigms. Processing continuous bias audits on edge-deployed ML models demands parallel computing pipelines that may exceed current edge server capacities.

Second, the efficacy of dynamic transparency mechanisms (pillar 2) relies heavily on the maturity of municipal smart city infrastructure. The proposition of broadcasting real-time ISAC activation states to public registries assumes the widespread deployment of interoperable urban Digital Twins. At present, such integrated civic data ecosystems remain largely theoretical or highly fragmented in practice. Translating this conceptual transparency into actionable civic interfaces requires overcoming significant data standardization deficits and establishing secure telemetry channels between private telecommunications operators and public municipal authorities.

Consequently, operationalizing trustworthy ISAC necessitates a robust, interdisciplinary research agenda. Future technical research must prioritize the development of lightweight algorithmic auditing protocols that do not degrade edge computing performance. Additionally, investigations into zero-knowledge cryptographic proofs for verifying hardware dormancy could allow operators to mathematically prove to regulators that sensing modules are inactive without exposing underlying network configurations. Concurrently, legal and standards bodies, such as 3GPP and ETSI, must formalize standardized API protocols that translate GDPR and AI Act compliance thresholds into machine-readable network configurations. Finally, realizing the full potential of 6G ISAC requires sustained collaboration between network architects, legal scholars, and urban planners to ensure the infrastructure is compliant by design prior to commercial deployment.

## V. Conclusion

The standardization and subsequent deployment of 6G Integrated Sensing and Communication (ISAC) introduces a paradigm shift in telecommunications, transitioning cellular networks into continuous environmental and biometric sensing grids. If deployed under an “always-on” architectural model, this ubiquitous device-free sensing fundamentally conflicts with the data minimization, purpose limitation, and biometric categorization mandates of the GDPR and AI Act. Cryptographic security alone cannot retroactively legitimize the non-consensual generation of spatial and behavioral intelligence.

To reconcile this structural collision, this paper proposes a sociotechnical governance framework that mandates regulatory compliance by design. By implementing three foundational pillars (purpose-bound hardware activation, dynamic civic transparency registries, and algorithmic auditing), telecommunications operators can ensure that device-free sensing capabilities are strictly triggered for authenticated, legally cleared mandates. This proposed architecture structurally prevents the unlawful accumulation of sensitive ambient data and mitigates the severe operational liabilities associated with unauthorized data harvesting.

Ultimately, rendering 6G ISAC legally viable requires an immediate, interdisciplinary intervention. Researchers across telecommunications engineering and digital law, alongside regulatory bodies, must collaboratively co-design these safeguards within current standardization frameworks, such as 3GPP and ETSI. All stakeholders involved in 6G standardization must fundamentally redefine the concept of "trustworthiness" beyond purely technical security metrics, systematically engineering it as strict sociotechnical and legal compliance before global hardware architectures are irrevocably finalized.

## ACKNOWLEDGMENT

This work was supported by the SweWIN center under Grant 2023-00572, funded by Vinnova, Sweden's Innovation Agency, and by Digital Futures.